\documentclass[letterpaper, 10 pt, conference]{ieeeconf}

\IEEEoverridecommandlockouts

\usepackage{amsmath}
\usepackage[ansinew]{inputenc}
\usepackage{graphicx}
\usepackage{subfig}
\usepackage{todonotes}

\title{\LARGE \bf
Acquire Driving Scenarios Efficiently: A Framework for Prospective Assessment of Cost-Optimal Scenario Acquisition
}

\author{Christoph Glasmacher$^{1}$, Michael Schuldes$^{1}$,  Hendrik Weber$^{2}$, Nicolas Wagener$^{1}$ and Lutz Eckstein$^{1}$
\thanks{$^{1}$The authors are with the Institute for Automotive Engineering, RWTH Aachen University, 52074 Aachen, Germany {\tt\small christoph.glasmacher@ika.rwth-aachen.de}}%
\thanks{$^{2}$The author is with the fka GmbH, 52074 Aachen, Germany}
}

\usepackage{tikz}
\usepackage{textcomp}
\usepackage{hyperref}
\usepackage{lipsum}

\newcommand\copyrighttext{%
	\footnotesize \textcopyright 2023 IEEE. Personal use of this material is permitted.
	Permission from IEEE must be obtained for all other uses, in any current or future 
	media, including reprinting/republishing this material for advertising or promotional 
	purposes, creating new collective works, for resale or redistribution to servers or 
	lists, or reuse of any copyrighted component of this work in other works. 
}
\newcommand\copyrightnotice{%
	\begin{tikzpicture}[remember picture,overlay]
	\node[anchor=south,yshift=10pt] at (current page.south) {\parbox{\dimexpr\textwidth-\fboxsep-\fboxrule\relax}{\copyrighttext}};
	\end{tikzpicture}%
}

\begin{document}

\maketitle
\thispagestyle{empty}
\pagestyle{empty}

\copyrightnotice

%%%%%%%%%%%%%%%%%%%%%%%%%%%%%%%%%%%%%%%%%%%%%%%%%%%%%%%%%%%%%%%%%%%%%%%%%%%%%%%%
\begin{abstract}

Scenario-based testing is becoming increasingly important in safety assurance for automated driving. However, comprehensive and sufficiently complete coverage of the scenario space requires significant effort and resources if using only real-world data. To address this issue, driving scenario generation methods are developed and used more frequently, but the benefit of substituting generated data for real-world data has not yet been quantified. Additionally, the coverage of a set of concrete scenarios within a given logical scenario space has not been predicted yet.
This paper proposes a methodology to quantify the cost-optimal usage of scenario generation approaches to reach a certainly complete scenario space coverage under given quality constraints and parametrization. Therefore, individual process steps for scenario generation and usage are investigated and evaluated using a meta model for the abstraction of knowledge-based and data-driven methods. Furthermore, a methodology is proposed to fit the meta model including the prediction of reachable complete coverage, quality criteria, and costs.
Finally, the paper exemplary examines the suitability of a hybrid generation model under technical, economical, and quality constraints in comparison to different real-world scenario mining methods.

\end{abstract}

%%%%%%%%%%%%%%%%%%%%%%%%%%%%%%%%%%%%%%%%%%%%%%%%%%%%%%%%%%%%%%%%%%%%%%%%%%%%%%%%
\section{INTRODUCTION}
Assuring the safety of automated vehicles is a key issue in current research.  
Thereby, scenario-based methods are a promising approach.
The required scenarios for the tests are mined from real data or created with generation methods \cite{Riedmaier.2020}.
Direct mining of scenarios from real recordings has the advantage that scenarios obtained in this way can be realistic consequently and can be traced back to a recording. However, this is associated with significant costs and efforts if the parameter space of a logical scenario shall be covered sufficiently complete. Therefore, multiple generation methods are developed that create concrete scenarios knowledge-based or data-driven. While these methods are less expensive than scenario mining from real-world data, they mostly lack the direct link to real traffic and thus the proof that the scenarios are realistic \cite{Cai.2022}.

Although scenario generation methods are increasingly used, a quantified proof of the economic viability of using these models is still lacking to the authors' knowledge. Therefore, a methodology is presented within this paper that allows a comparison between generation methods and real-world scenario mining methods as well as a combination of the two approaches to find cost-optimal combinations.
The used comparison objective is thereby the coverage of a predefined scenario space for a set of concrete scenarios and the necessary cost to achieve a sufficient degree of coverage within this space. Since models can differ in their characteristics, quality constraints are set.

In order to examine to what extent scenario generation methods can substitute the recording of real-world data, a meta model is set up to make different methods comparable. This can be used to evaluate the benefit of mining and generation approaches from a technical and economic point of view. Thereby, the quality of a scenario generation approach is measured by the coverage of the scenario space and the model error depending on the number of generated scenarios. A method is presented for deriving the parameters of the meta model from an exemplary generation model. This filled meta model is evaluated with an economic cost analysis according to incurred costs to reach a certain coverage. This method and the assumptions made are tested using a multivariate generation model \cite{ChristophGlasmacherHendrikWeberMichaelSchuldesNicolasWagenerLutzEckstein.2023} with real-world data from the inD dataset \cite{Bock.2020}.
Finally, sensitivity analyses on the influence of different factors as quality requirements and costs are assessed to consider the relevance of generation models under certain conditions.

\section{RELATED WORK}

Scenario mining and generation approaches became relevant as an input for scenario-based testing of highly automated driving functions. Thereby, different definitions of scenarios are published. A common definition is that a scenario is a sequence of scenes and can be described by actions and triggers \cite{ISO.}.

\subsection{Scenario Mining and Generation}

A distinction can be made between the direct mining of scenarios from real traffic and model-based scenario generation.
The direct mining from real-world data is possible using different sources effecting technical attributes and costs to acquire these scenarios.
\cite{Kessler.2012} extracts real traffic data using sensors on a vehicle, while traffic data in \cite{Kloker.} is recorded using infrastructure sensors leading to a more comprehensive detection of the scenario. A similar approach is taken by \cite{Bock.2020}, as in this case, trajectories can be generated via drone recordings.

In addition to the creation of concrete scenarios from real-world data, these can be created using generation methods. A distinction is made between knowledge-based and data-driven methods \cite{Cai.2022}.
Knowledge-based methods rely purely on expert knowledge that can be described, for example, by ontologies \cite{Bagschik.2022}. 
Data-driven methods such as \cite{Gelder.2022}, on the other hand, are based on the extraction of parameters and distribution functions from real data.
Hybrid methods combine knowledge-based and data-driven approaches by using real distributions and knowledge-based dependencies \cite{ChristophGlasmacherHendrikWeberMichaelSchuldesNicolasWagenerLutzEckstein.2023}.

\subsection{Scenario Space Coverage}
\label{sec:scenario_space_coverage}

Both in development and safety assurance of automated driving functions the coverage of a scenario space is an important measure.
Thereby, the scenario abstraction layers concrete, logical, abstract and functional can be distinguished \cite{Neurohr.2021}.
Proposed methods are either focusing on extracting concrete scenarios to reach a locally sufficient coverage or on arguing coverage of scenario concepts. \cite{Weber.2023} creates an ontology to cover abstract traffic scenarios and derive parameters from those. \cite{Vater.2021} discretizes a given parameter space into locally equally distributed buckets wherein a scenario should be assessed to sufficiently test a limited parameter space. \cite{Weng.2022} defines a delta room for this so that no buckets are needed and the space can be covered continuously.  
Another sampling method is proposed by \cite{James.2017} who uses a bootstrapping approach.

\subsection{Cost Analysis}

The cost reduction and causation of costs of automated driving is discussed from different perspectives. \cite{Wadud.2016} deals with the benefits and possible cost reductions while \cite{.2015} refers to indirect costs such as insurance.
Although scenario-based testing is also qualitatively motivated by effort reduction \cite{Galbas.2022}, quantitative analysis within scenario-based testing has played a minor role so far.
E.g. \cite{.2019} looks at the economic feasibility of simulating test scenarios by estimating costs only caused by simulation.

In other areas, cost analyses are considered more intensively and thus process chains are broken down to assign costs to each step. For example, \cite{Tosello.2019} breaks down a cost analysis based on the additive manufacturing process chain with itemized costs.

\section{METHODOLOGY}

To compare costs of different scenario generation and mining approaches, three aspects are highlighted: The objective of different scenario acquisition methodologies has to be the same to ensure comparability. This is derived from Sec.~\ref{sec:scenario_space_coverage}. The premise is set as the achievement of a sufficient coverage of a given scenario parameter space by generating concrete scenarios. Therefore, the coverage of the given space has to be approximated to calculate the costs referring to the amount of individual scenarios to be generated. 
Furthermore, scenario acquisition approaches are transferred into a meta model. Using a common scenario acquisition meta model allows easier comparability and the combination of models to evaluate occuring costs and technical specifications against single approaches.
Both, meta model and coverage approximation, are then used as input for the economic analysis wherein the different cost attributes are compared to evaluate the cost-optimal strategy for reaching a sufficient coverage under given quality constraints.

\subsection{Coverage Approximation}
\label{sec:completeness_approximation}

\begin{figure}[tb]
	\vspace{0.1cm}
	\centering
	\includegraphics[width=\linewidth]{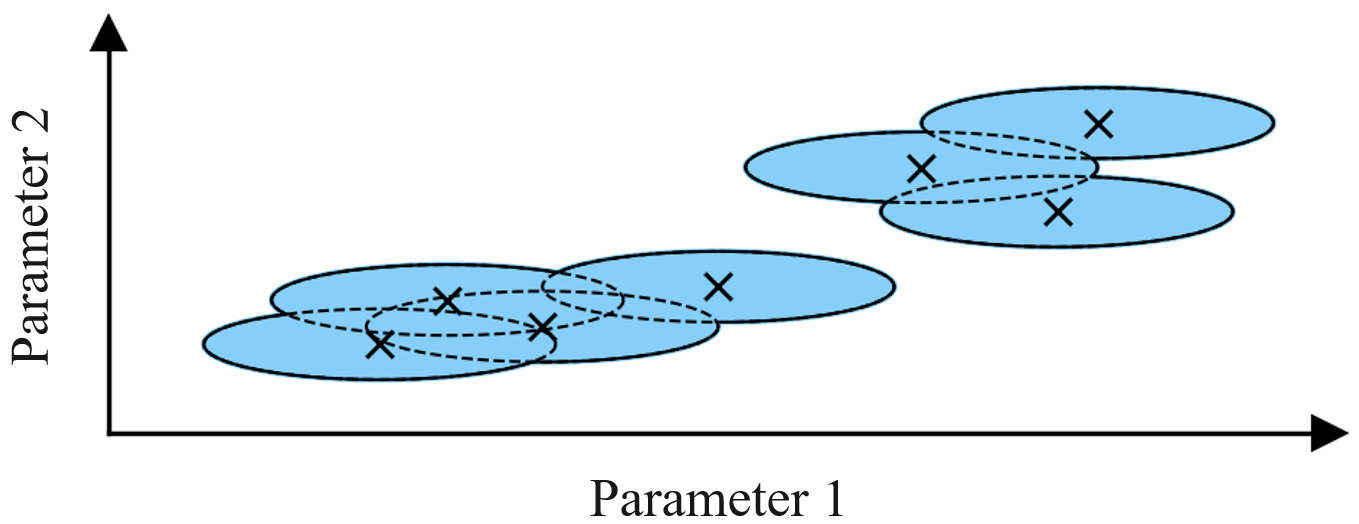}
	\caption{Exemplary coverage of parameter space}
	\label{figure:completeness_area}
	\vspace{-0.1cm}
\end{figure}

As shown in the state of the art (see Sec.~\ref{sec:scenario_space_coverage}), the coverage of a given scenario space is not predicted a priori but sampled within given ranges. 
Since it is infeasible to perform an economic analysis to help make a decision before knowing the actual ranges, we approximate the coverage of the scenario space. Therefore, a scenario is represented by a point in the high-dimensional and partly continuous parameter space. This space is considered to be covered by a set of scenarios if the local density is higher than a minimum density. For this purpose, an ellipsoid ($E_i$) is created around each parameter point $\hat{x} = {\hat{x}_1, ... \hat{x}_m}$ (Eq.~\ref{eq:ellipsoid}) defining the $m$-dimensional space in which no other parameter needs to be set to reach the coverage criteria (see Fig.~\ref{figure:completeness_area}). Thereby, the ellipsoid can be stretched and compressed in each dimension ($p_j$) according to requirements. The volume of the unification of all ($n$) scenario ellipsoids is used to determine the overall coverage of the scenario space ($V$) % (Eq.~\ref{eq:completeness_union},
Eq.~\ref{eq:volume_completeness}). Instead of discretizing the space a priori, the proposed method gives the opportunity to calculate the density more accurately without discretization limits.

\begin{equation}
	E_i(x) = \begin{cases}
	1, & 1 \geq \sum_{j=1}^{m} (\frac{x_j - \hat{x_j}}{p_j})^2 \\
	0, & else
	\end{cases}
	\label{eq:ellipsoid}
\end{equation}

\begin{equation}
V = \int \bigcup_{i}^{n} E_i(x) dx
\label{eq:volume_completeness}
\end{equation}

To estimate the achieved degree of approximated coverage of the predefined scenario parameter space, the unified area of the body $V$ is calculated. For a stochastic distribution of the input data, the area can be represented by a cumulative Weibull distribution function \cite{Guthrie.2020} for a sufficiently large number of scenario input data ($x$) and an offset of initial coverage from input data $V_{pre}$ (Eq. \ref{eq:weibull}). 
Thereby, the observed parameter space can be restricted a priori depending on the requirements, so that, for example, the scenario coverage of the parameter space can be determined for a relevant sub-space like conflicts.
The degree of coverage ($C$) is thus described from the covered space ($V(x)$) in relation to the limit of the saturation function ($a$) and potential initial covered space ($V_{pre}$) (Eq.~\ref{eq:completeness_coefficient}).
\begin{equation}
V_{model}(x) = a \cdot (1 - e^{-b \cdot x^c}) + V_{pre}
\label{eq:weibull}
\end{equation}

\begin{equation}
C(x) = \frac{V(x)}{a + V_{pre}}
\label{eq:completeness_coefficient}
\end{equation}

\subsection{Scenario Acquisition Meta Model}
\label{sec:metamodel}

In order to compare different scenario generation and mining models, a meta model is set up. It condenses scenario acquisition to main attributes to reach an easier comparability and implementation in the framework. The model does not distinguish between the extraction of real-world data and the synthetic generation of scenarios. The core is the coverage calculation depending on the number of scenarios to be generated as well as the coverage already achieved with respect to an already existing data basis ($V_{pre}$). To describe this relation, an individual coverage approximation function has to be set. This can differ depending on the acquisition method and sampling strategy. For a stochastically distributed sampling, a Weibull function can be used as in Sec.~\ref{sec:completeness_approximation}. However, the function is limited by a straight line dependent on the coverage of an individual scenario (Eq.~\ref{eq:ellipsoid}).
Besides the parameters describing the coverage, quality criteria and arising costs are considered. To describe the quality criteria, quality requirements are imposed on the model in the form of an allowed error rate. This is needed since the quality of the generated or mined scenarios can diverge.

A third column of the meta model includes a categorization of incurring cost for the scenario generation or mining. Thereby, three types of costs are distinguished:

\begin{itemize}
	\item Setup cost: Cost for setting up the generation model or sensors and toolchains to record data
	\item Gaining cost: Average cost for generating or mining of a scenario
	\item Validation cost: Average cost for proofing if a scenario is valid or invalid 
\end{itemize}
The described parameters vary strongly depending on the scenario approach. While an error rate for manually labeled scenarios should be close to zero, costs for setup and data mining are significantly higher than for a knowledge-based scenario model.

\begin{figure}[tb]
	\vspace{0.15cm}
	\centering
	\includegraphics[width=\linewidth]{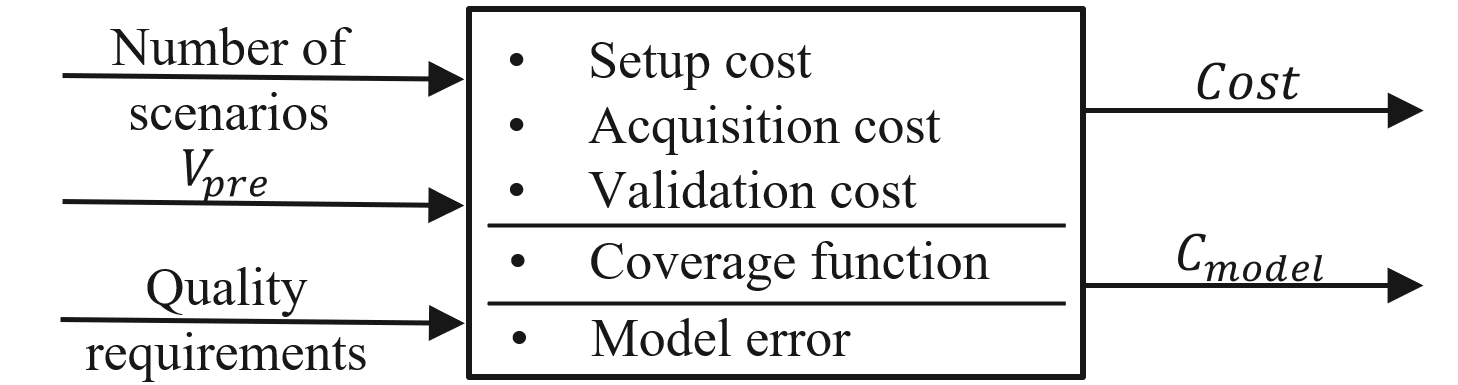}
	\caption{Scenario acquisition meta model}
	\label{figure:meta_model}
	\vspace{-0.15cm}
\end{figure}

\subsection{Meta Model Fitting}
\label{sec:metamodel_fitting}

Since scenario acquisition methods can have fundamentally different characteristics, meta models and their attributes have to be setup for each method individually. 
While the economic attributes of the meta model are dependent on various influencing factors and must be extracted from expert knowledge, technical factors as coverage approximation and model error can be determined by experiments.

The coverage of a predefined scenario parameter space is approximated with newly acquired scenario parameter sets from the investigated acquisition method.
These scenarios are used to approximate the coverage according to Sec.~\ref{sec:completeness_approximation} with a bootstrapping approach.
Thereby, only valid scenarios are considered for the calculation.
 
To detect invalid scenarios, a sufficiently large database of valid scenarios is needed. The covered scenario space (Eq.~\ref{eq:volume_completeness}) and approximated coverage threshold (Eq.~\ref{eq:completeness_coefficient}) are calculated for this input. To approximate a potentially complete coverage according to the approximated threshold, the given volume is thickened under logical constraints.
A scenario is then considered valid if its parameter representation $p_i$ is in the approximated volume ($V_{comp}$). So, a general model error rate ($e_{model}$) can be calculated (Eq.~\ref{eq:model_error_calculation}) from a sufficiently large amount of generated parameter sets ($n$).

\begin{equation}
e_{model} = \frac{\sum_{i=1}^{n}\Bigg\{
	\begin{split}
	1, & ~~~~ p_i \subset V_{comp}\\
	0, & ~~~~ else
	\end{split}}{n}
\label{eq:model_error_calculation}
\end{equation}

\subsection{Cost analysis}
\begin{figure*}[tb]
	\vspace{0.15cm}
	\centering
	\includegraphics[width=\linewidth]{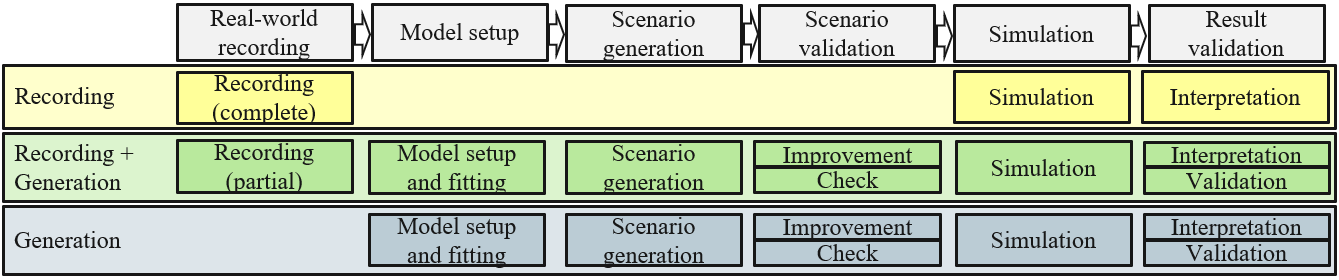}
	\caption{Cost flow model for different types of scenario acquisition}
	\label{figure:cost_flow}
	\vspace{-0.15cm}
\end{figure*}

Based on the derived meta models, respective costs are analyzed. According to meta model characteristics, costs are incurred differently along the utilization chain of scenarios differently. Thereby, the complete process chain from scenario acquisition to interpretation of possible simulations is analyzed (see fig. \ref{figure:cost_flow}).
These costs depend on different influencing factors depending on a combination of personal costs, licenses, tools and more. Those are summed up for a comprehensive evaluation %(Eq.~\ref{eq:cost_commulation}) 
and detail the categories stated in the meta model (sec.~\ref{sec:metamodel}).%:

While most costs are fixed or solely dependent on the number of generated scenarios, validation costs of the scenarios are dependent on quality criteria and model characteristics. The costs are compounded of the costs for validation and the costs for dataset improvement after generation. Improvement may be necessary if the quality requirements of the stakeholders cannot be met or a potential quality analysis may become infeasible large. Within the improvement step the initial error rate of the model ($e_{model, initial}$) can be decreased ($e_{model,optimized}$) by checking scenarios ($n_{improv}$) and sorting out invalid scenarios (Eq.~\ref{eq:upfront_check}).
\begin{equation}
n_{improv} = \max \Big(\Big\lceil 1 - \frac{e_{model,optimized}}{e_{model, initial}} \Big\rceil, 0\Big)
\label{eq:upfront_check}
% maybe nicer to do with cases
\end{equation}

To check whether the error rate of a acquired scenario set is below an allowed error rate, a randomly selected and representative sample ($n_{rand}$) is investigated using the Cochran formula (Eq.~\ref{eq:spot_check}) \cite{Cochran.2005}. Thereby, randomly taken samples are checked for its validity to reach a certain confidence for the complete dataset. Thereby, the process of checking is not further specified but should be reflected in checking costs. The needed sample size to reach a certain confidence ($Z$) is dependent on the expected error of the model ($e_{model}$), and the allowed tolerance ($e_{tol}$) of the estimate (Eq.~\ref{eq:spot_check}). The allowed tolerance depends on the difference between the allowed and actual model error to be sure not to exceed the allowed error rate. So, an optimization within the improvement step may be reasonable although the error rate may be below the allowed rate.

\begin{equation}
n_{rand} = \bigg\lceil \frac{Z^2 \cdot e_{model} \cdot (1-e_{model})}{e_{tol}^2} \bigg\rceil
\label{eq:spot_check}
\end{equation}

However, since the improvement step may have already tested some of the scenarios that could be randomly selected again in the sample, the sample size for the randomized spot check that is still to be tested ($n_{rand, corr}$) can be reduced by this amount (Eq.~\ref{eq:reduced_spot_check}).
Thereby, due to different test sizes for data improvement and validation, an optimization problem arises (Eq.~\ref{eq:overall_check_size}) to calculate the cost-optimal error rate to reach after improvement ($e_{opt}$) and the optimal check size ($n_{check}$).

\begin{equation}
	n_{rand, corr} = \max \Big(\Big\lceil n_{rand} - n_{improv} \cdot \frac{n_{rand}}{n} \Big\rceil, 0\Big)
	\label{eq:reduced_spot_check}
\end{equation}

\begin{equation}
n_{check} = (n_{improv} + n_{rand, corr}) \big\vert_{\frac{dn}{de_{opt}}=0}
\label{eq:overall_check_size}
\end{equation}

\section{RESULTS}
\label{sec:results}

For the application of the proposed framework, the individual steps are performed and evaluated sequentially. The coverage assumptions are checked and a multivariate data-driven scenario generation model is fitted to the meta model. Those are used as inputs for a sensitivity analysis to investigate the influencing cost factors for real-world data, necessary quality criteria, and completeness requirements. 
The inD dataset \cite{Bock.2020} is used as a data basis.

For the sensitivity analysis, costs for model setup and scenario generation are assumed to be relatively small compared to real-world recording and validation. Furthermore, costs for simulation and result interpretation are negligible for a comparison since similar costs arise for each method. The costs for result validation may be relevant in case of insufficient quality at simulation onset but are strongly dependent on the use case. So, it is assumed that the scenario error for simulation input is sufficiently small after checking to neglect further result validations.

\subsection{Completeness Coverage Approximation}
\label{sec:results_completeness_approximation}

\begin{figure}[tb]
	\centering
	\subfloat[Occupied area (blue), saturation function (yellow) and approximated threshold (red)]{
		\includegraphics[width=0.45\linewidth]{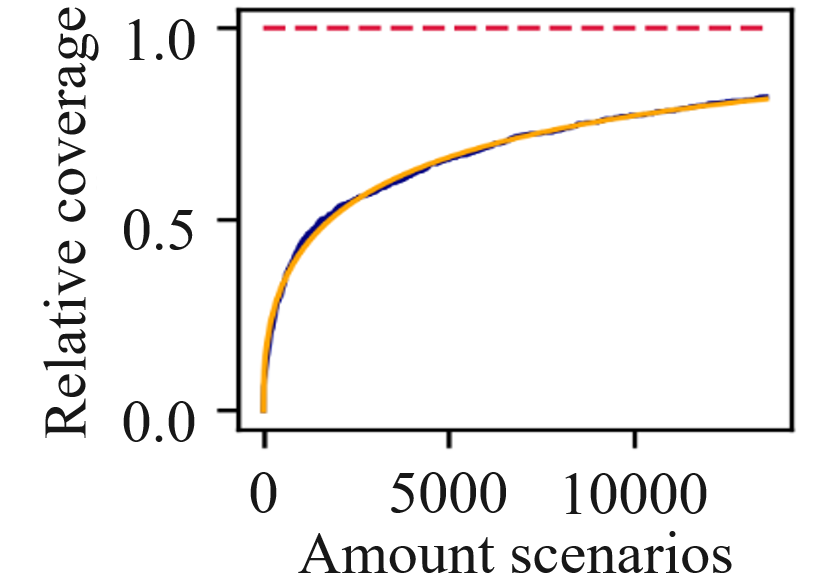}
		\label{fig:completeness_fit_area}
	}
	\hfill
	%\vspace{0.03cm}
	\subfloat[Weibull parameters]{ 				  
		\includegraphics[width=0.45\linewidth]{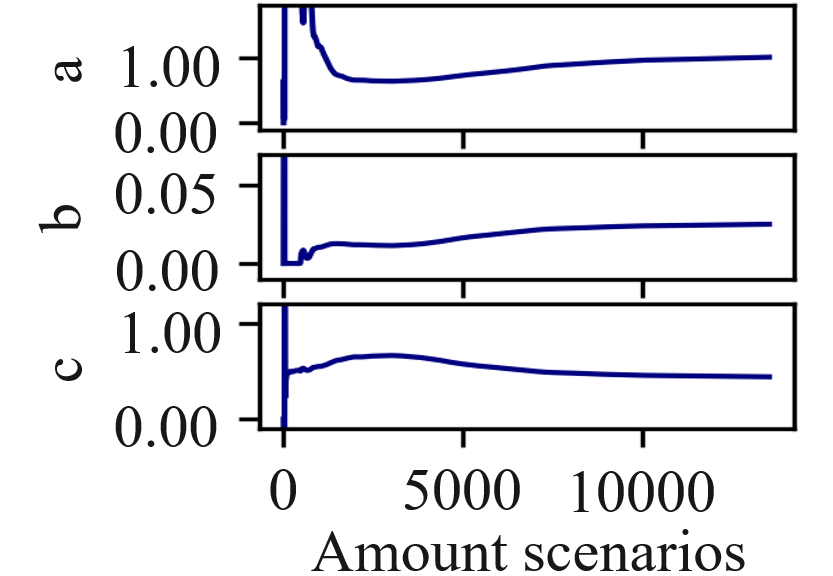}
		\label{fig:completeness_fit_parameters}
	}
	\caption{Real-world data coverage}
	\label{fig:completeness_real_data}
\end{figure}

To determine the coverage of real-world data within the dataset, two-dimensional parameter sets are used as scenario representation. Therefore, input and output speeds of the road users are chosen and for each scenario combined to two-dimensional parameter sets. The 13,599 extracted parameter sets are used in randomized order to improve fitting results  by neglecting recorded patterns.
Joining the predefined volume (Eq.~\ref{eq:ellipsoid}) using a bootstrapping approach leads to converging parameters of the coverage approximation function for increasing amount of input scenarios (see Fig.~\ref{fig:completeness_real_data}). This shows that the coverage approximation can be estimated with a cumulative Weibull distribution function. Thus, the completeness coverage approximation of the dataset with given parameters is 81.52 percent.
Performing this fitting with different sets of individual scenarios the Weibull function fits the coverage of real-world data better if the coverage of an individual scenario remains small since scenarios can only be added discretely. If the coverage of a scenario is relatively large, the fitted function overestimates the coverage and completeness.

Additionally to the inD dataset, an internal dataset containing around 1.6 million trajectories with limited invalid scenarios is assessed. Due to the broader spreading of errors and their seldom occurrence, fitting the complete dataset within one function does not lead to a convergence of the Weibull parameters. The parameter for the limit of the coverage increases slightly because the slope rate is lower than the rate of valid values. So, the parameters can be interpreted as superposed functions for valid and invalid scenarios %.
and error rates for outliers can be estimated.

\subsection{Generation Model Fitting}
\label{sec:results_meta_model_fitting}

\begin{figure}[tb]
	\centering
	\subfloat[Model error for different amounts of real-world input values]{ 				  
		\includegraphics[width=0.45\linewidth]{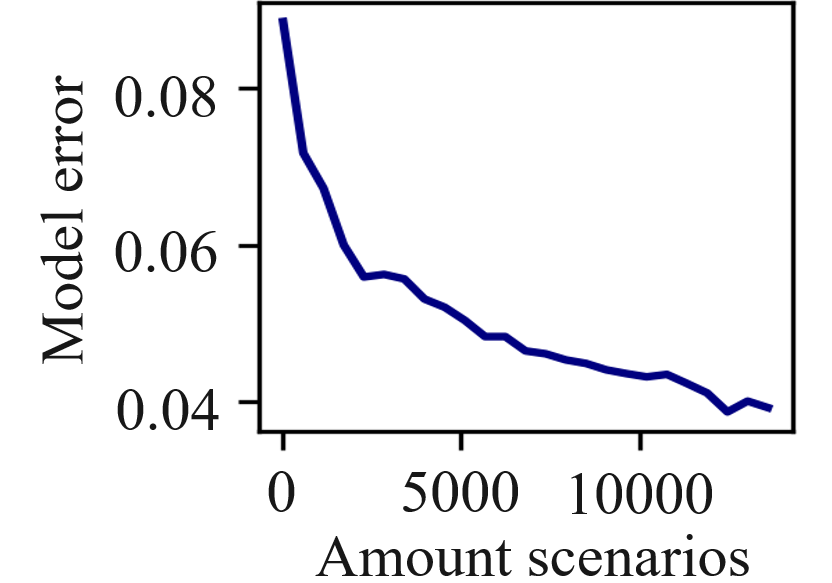}
		\label{fig:model_error_over_input}
	}
	\hfill
	\subfloat[Coverage functions (blue, orange) for different entry points with approximated threshold (red)]{
		\includegraphics[width=0.45\linewidth]{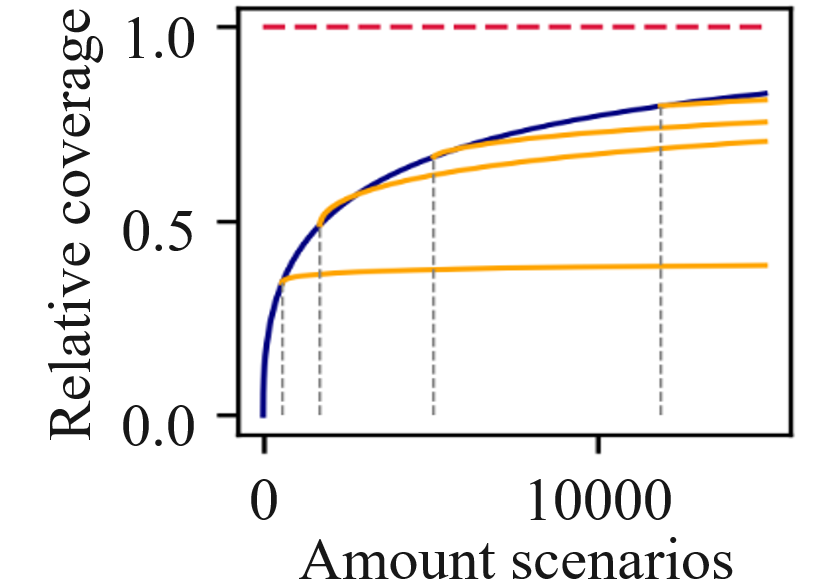}
		\label{fig:completeness_over_input}
	}
	\caption{Fitting of meta model}
	\label{fig:meta_model_fitting}
\end{figure}

Based on the completeness analysis of the real-world data, a multivariate generation model is fitted using different amounts of input data.  
In order to determine coverage parameters and model errors, 50,000 valid two-dimensional parameter sets are generated for each amount of input data. The check takes place by the comparison with an approximated complete volume (see Sec.~\ref{sec:results_completeness_approximation}) %.
according to Sec.~\ref{sec:metamodel_fitting}.
Evaluating the meta model parameters shows that the error decreases as the number of input data increases. The achievable combined coverage from real data mining and generation increases with the increasing number of real-world data (see Fig.~\ref{fig:completeness_over_input}). Since the sampling is done randomly, the slopes are below the real-world data. A more systematic sampling could increase those slopes but would likely also increase the model error due to less scenarios within a confident region.

\subsection{Analysis of Combined Approaches}

Data-driven scenario generation models as parametrized in Sec.~\ref{sec:results_meta_model_fitting} need a data basis to be fitted. The data has to be mined a priori. Towards an overall sufficiently complete coverage, the costs of both data mining and further generation have to be combined. The sweet spot until when data should be mined and when the generation should start is an optimization problem influenced by technical, quality and economic factors. One significant factor is the influence of cost for real-world mined scenarios since costs could differ significantly. In order to only have valid scenarios for real-world mining, those have to be labeled. The costs are depending on sensors and suppliers so that exemplary values are set by expert knowledge based on labeling effort for a scenario for drone (90\,€), infrastructure (240\,€) and onboard data (740\,€).
Thereby, a scenario is assumed to be labeled for 10 seconds and to be sampled with 10 Hertz.

\begin{figure}[tb]
	\vspace{0.1cm}
	\centering
	\includegraphics[width=\linewidth]{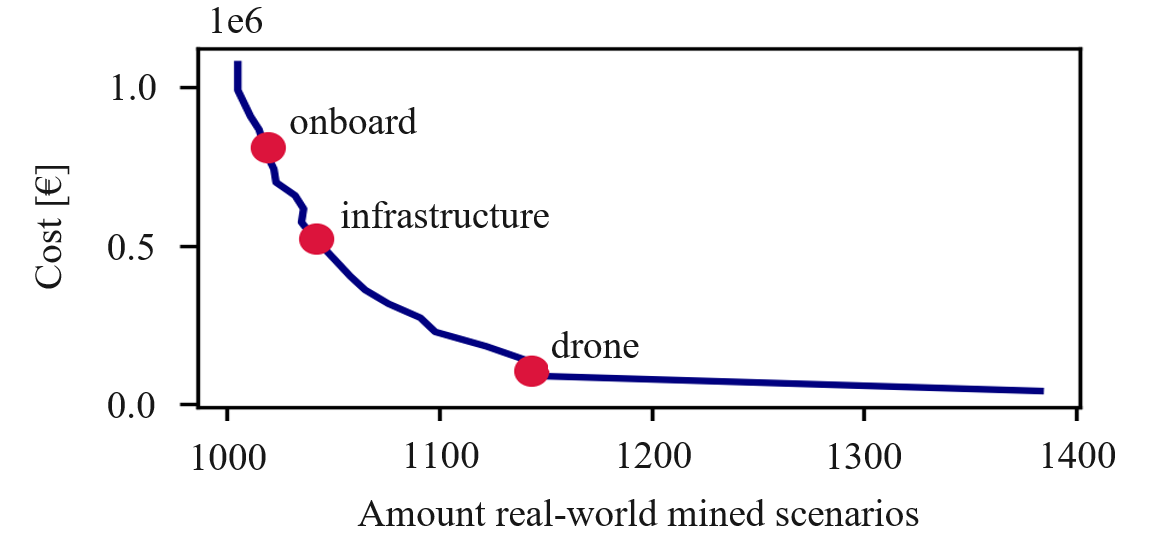}
	\caption{Cost-optimal combination of scenario acquisition based on expert-based scenario mining costs}
	\label{figure:sensitivity_infrastructure}
	\vspace{-0.1cm}
\end{figure}

Varying scenario mining costs lead to a shift in the number of scenarios to be generated. The overall cost do not correlate linearly with the costs for scenario mining but show a logarithmic influence (see Fig.~\ref{figure:sensitivity_infrastructure}). This behavior is caused by three factors: Besides the reduction in scenario mining costs, the quality of the generated scenarios is improved so that checking costs decrease. Furthermore, the slope for achieving completeness increases so that less additional scenarios have to be generated.
When varying the scenario generation entry point, overall costs increase or the criteria cannot be met. Increasing the number of mined scenarios will still meet the criteria if it can be assumed that they are valid. Still, costs would be increased. Using less mined scenarios increases costs due to more and qualitatively worse generated scenarios. Furthermore, at a certain point, the requested coverage cannot be reached anymore according to Fig.~\ref{fig:completeness_over_input}.

\subsection{Influence of Quality Requirements}

\begin{figure}[tb]
	\centering
	\includegraphics[width=\linewidth]{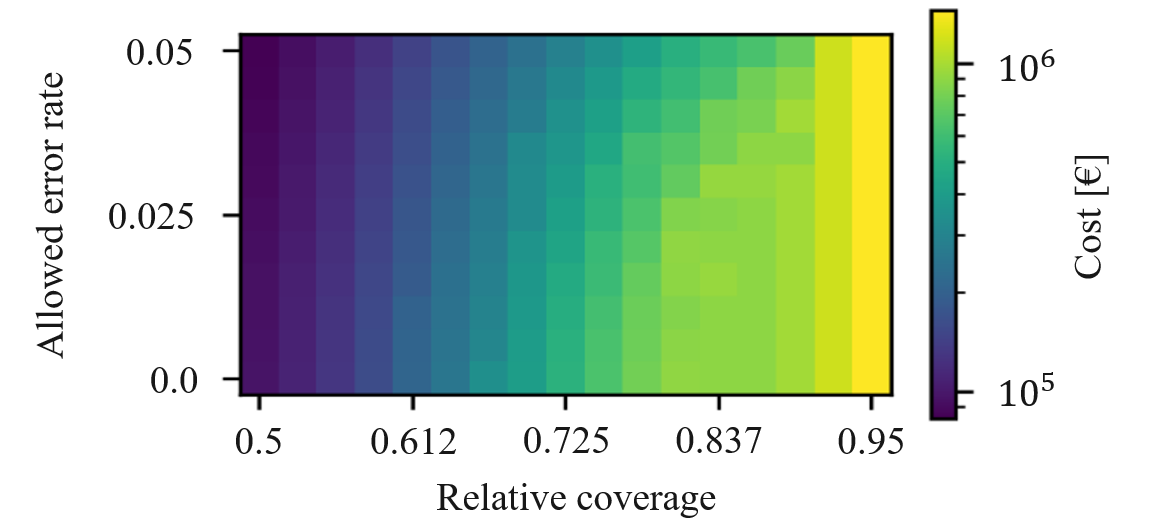}
	\caption{Quality requirement costs influence}
	\label{figure:sensitivity_quality}
\end{figure}

Quality requirements and data verification costs are significant cost factors for error-prone generation methods.
Since multiple factors are driving those costs, the allowed error rate in the final dataset and the completeness coverage requirements are analyzed exemplary. Thereby, the minimal costs according to the generation entry point are calculated (see Fig.~\ref{figure:sensitivity_quality}).
Depending on the verification costs for a scenario, it can be seen that the completeness criteria has an exponential and larger influence than costs to decrease the scenario error. Those are related to scenario checking costs. Its effect decreases and turns around when costs for checking become similarly expensive as scenario mining. Thereby, the influence of the allowed error rate is highly dependent on the quality of the used generation model. The better the model the lower is the influence since nearly no improvement has to be made and spot check costs remain small.

\section{DISCUSSION}

The exemplary application of the framework shows that it can give guidance to use different methods to reach a cost-optimal and desired scenario space coverage.
By investigating the inD~dataset, it can be seen that relative coverage can be approximated by using bootstrapping. Therefore, a reasonable amount of input data is needed to reach a convergence depending on the spreading and dimensionality of parameters. This assumption of completeness, thereby, holds under a given parameter space and bias of input data. For example, if only daytime driving is recorded, completeness can only be approximated for daytime driving under given parametrization.
However, insofar data are subject to errors, convergence of the real-world data is no longer necessarily achievable but the approximated coverage could increase slightly with outliers. Therefore, convergence of coverage parameters can be used as a quality measure and error detection mechanism.
Besides focusing purely on coverage, it would also be possible to extend the method to distributions to make statements about representativeness of individual scenarios within an approximated complete parameter space.

Within the economic benefit analysis, it is shown that the economic feasibility substituting real-world data can be calculated using the proposed meta model. The feasibility highly depends on the required quality and various cost factors, as well as the technical aspects of the generation model. So, it must be evaluated for the individual case. However, it can be shown that a partial substitution is feasible for reasonable requirements under realistic conditions. Nevertheless, under investigated circumstances, real-world data cannot be substituted completely.

Beyond these monetary expressible factors, further aspects could be considered within an evaluation. For example, a realistic exploration of the scenario space leading to a parameter space is limited without the use of sufficient real-world data. 
In addition, factors such as unknown scenarios, additional knowledge gain about real traffic as well as the greater emotional trust in real-world data are not quantified but could partly be added to the model. For quantification of soft factors as e.g. social acceptance, surveys could be executed.

\section{CONCLUSIONS}
This paper presents a framework for cost-optimal scenario acquisition. Thereby, it allows a comparison of different scenario generation models as well as a comparison with real-world scenario mining methods. The scenario acquisition objective is thereby set as a certainly complete coverage of a scenario space within a logical scenario. It is shown that a stochastic input of scenario data is represented by a saturation function so that the achievement of a desired degree of coverage can be approximated within a given logical scenario space. Based on this, a methodology is introduced to transform scenario mining and generation models into a common meta model. Those are used as inputs for the cost analysis. The analysis allows a quantified evaluation of generation costs and gives guidance on a methodology or combination of methodologies to choose a cost-optimal scenario acquisition.
We show that generation models can replace real-world data mining under certain quality and technical constraints economically feasible.
Beyond the econometric considerations, a quantification of additional characteristics remains future work.

\addtolength{\textheight}{-12cm}

%%%%%%%%%%%%%%%%%%%%%%%%%%%%%%%%%%%%%%%%%%%%%%%%%%%%%%%%%%%%%%%%%%%%%%%%%%%%%%%%

%%%%%%%%%%%%%%%%%%%%%%%%%%%%%%%%%%%%%%%%%%%%%%%%%%%%%%%%%%%%%%%%%%%%%%%%%%%%%%%%

%%%%%%%%%%%%%%%%%%%%%%%%%%%%%%%%%%%%%%%%%%%%%%%%%%%%%%%%%%%%%%%%%%%%%%%%%%%%%%%%

\section*{ACKNOWLEDGMENT}

The research leading to these results is funded by the German Federal Ministry for Economic Affairs and Climate
Action within the project "Verifikations- und Validierungsmethoden automatisierter Fahrzeuge im urbanen Umfeld".
The authors would like to thank the consortium for the successful cooperation.

%%%%%%%%%%%%%%%%%%%%%%%%%%%%%%%%%%%%%%%%%%%%%%%%%%%%%%%%%%%%%%%%%%%%%%%%%%%%%%%%

\bibliographystyle{IEEEtran}
\bibliography{bibtex/literature.bib}

\end{document}